\PassOptionsToPackage{kerning}{microtype}
\documentclass[%
  anonymous=false,%
  authordraft=false,%
  format=sigconf,%
  review=false,%
  screen=true,%
  timestamp=false,%
  pbalance,%
]{acmart}

\copyrightyear{2022}
\acmYear{2022}
\setcopyright{acmlicensed}
\acmConference[ICSE '22 Companion]{%
  44th International Conference on Software Engineering Companion%
}{May 21--29, 2022}{Pittsburgh, PA, USA}
\acmBooktitle{%
  44th International Conference on Software Engineering Companion %
  (ICSE '22 Companion), May 21--29, 2022, Pittsburgh, PA, USA%
}
\acmPrice{15.00}
\acmDOI{10.1145/3510454.3516829}
\acmISBN{978-1-4503-9223-5/22/05}

\usepackage{defs}
\graphicspath{{./}{./img/}}

\newcommand\numModules{\num{118}\xspace}
\newcommand\numIterations{\num[round-precision=2]{30}\xspace}
\newcommand\timeout{\SI{600}{\second}\xspace}
\newcommand\numProjects{\num[round-precision=2]{17}\xspace}
\newcommand\avgCoverage{\SI{67.96650716076552}{\percent}\xspace}
\newcommand\avgCoverageDynaMOSA{\SI{67.96650716076552}{\percent}\xspace}
\newcommand\avgCoverageMIO{\SI{66.99329239952593}{\percent}\xspace}
\newcommand\avgCoverageMOSA{\SI{67.78516671267488}{\percent}\xspace}
\newcommand\avgCoverageRandom{\SI{63.56828612377363}{\percent}\xspace}
\newcommand\avgCoverageWS{\SI{66.9088489056397}{\percent}\xspace}
\newcommand\avgCoverageWSA{\SI{67.53200462581992}{\percent}\xspace}

\begin{document}

\title{Pynguin: Automated Unit Test Generation for Python}

\author{Stephan Lukasczyk}
\email{stephan.lukasczyk@uni-passau.de}
\orcid{0000-0002-0092-3476}
\affiliation{%
  \institution{University of Passau}
  \city{Passau}
  \country{Germany}
}

\author{Gordon Fraser}
\email{gordon.fraser@uni-passau.de}
\orcid{0000-0002-4364-6595}
\affiliation{%
  \institution{University of Passau}
  \city{Passau}
  \country{Germany}
}


\begin{abstract}
  %
  Automated unit test generation
  is a well-known methodology aiming
  to reduce the developers' effort of writing tests manually.
  %
  Prior research
  focused mainly on statically typed programming languages like Java.
  In practice, however,
  dynamically typed languages have received a huge gain in popularity
  over the last decade.
  %
  This introduces the need for tools and research
  on test generation for these languages, too.
  %
  We introduce \pynguin,
  an extendable test-generation framework for Python,
  which generates regression tests with high code coverage.
  \Pynguin is designed to be easily usable by practitioners;
  it is also extensible
  to allow researchers to adapt it for their needs
  and to enable future research.
  We provide a demo of \pynguin at
  \url{https://youtu.be/UiGrG25Vts0};
  further information,
  documentation, the tool, and its source code are available at
  \url{https://www.pynguin.eu}.
\end{abstract}


\maketitle

\section{Introduction}\label{sec:introduction}

Automated software test generation has a long history
both in research
and industrial settings.
Over the years,
researchers have presented many approaches for generating
 test input data,
such as
random~\cite{DN84} and search-based~\cite{McM04} techniques.
Many of these approaches have been implemented in tools,
most notably \toolname{Randoop}~\cite{PLE+07}
and \toolname{EvoSuite}~\cite{FA11}
for the Java programming language.
%

The focus on the Java programming language, however, represents a limitation of
prior research, as dynamically typed programming languages such as JavaScript
and Python have gained huge popularity over the last decade.
Python,
in particular,
is popular in many domains such as data science and machine learning, and
it nowadays ranks as one of the most used programming languages~(see,
for example, the IEEE Spectrum ranking\footnote{%
  \url{https://spectrum.ieee.org/top-programming-languages/}, %
  last accessed 2022–02–10.%
}).
This increasing popularity of the language
requires more and better tools
to support the developers
and improve the general code quality they produce.
Testing is one of the most important techniques to improve the quality of software, but automated tool support for test generation
is currently lacking in the Python tool box.
Unfortunately, automated test generation is challenging
in the context of a dynamically typed language.
%

A crucial problem impeding the development of test generation techniques is
the fact that programs written in a dynamically typed language
usually do not provide any information about variable types.
Statically deriving this information is hard~\cite{GPS+15}, as these
languages often allow to change the type of a variable's value throughout the
program, dynamically modify objects at runtime, or provide type coercions that
might not match the intent of the programmer~\cite{PS15}.
As a consequence, test generation research has so far only tackled specific aspects or restricted scenarios. For example,
\toolname{TSTL}~\cite{HGP+18},
\toolname{CrossHair}\footnote{%
  \url{https://crosshair.readthedocs.io},
  last accessed 2022–02–10.
},
\toolname{Klara}\footnote{%
  \url{https://klara-py.readthedocs.io},
  last accessed 2022–02–10.
}, or
\toolname{Auger}\footnote{%
  \url{https://github.com/laffra/auger},
  last accessed 2022-02–10.
},
 require manual effort by the developer
before they can produce test cases.
Although \toolname{Hypothesis}~\cite{MH19} can also generate tests automatically, its aim is to find minimal property-violating input values.
We therefore introduce \pynguin~\cite{LKF20}, which aims to use search-based test-generation techniques to automatically generate regression tests~\cite{Xie06} with high code coverage, without requiring user input.
%
%

\begin{figure*}
  \centering
  \includegraphics[width=\textwidth]{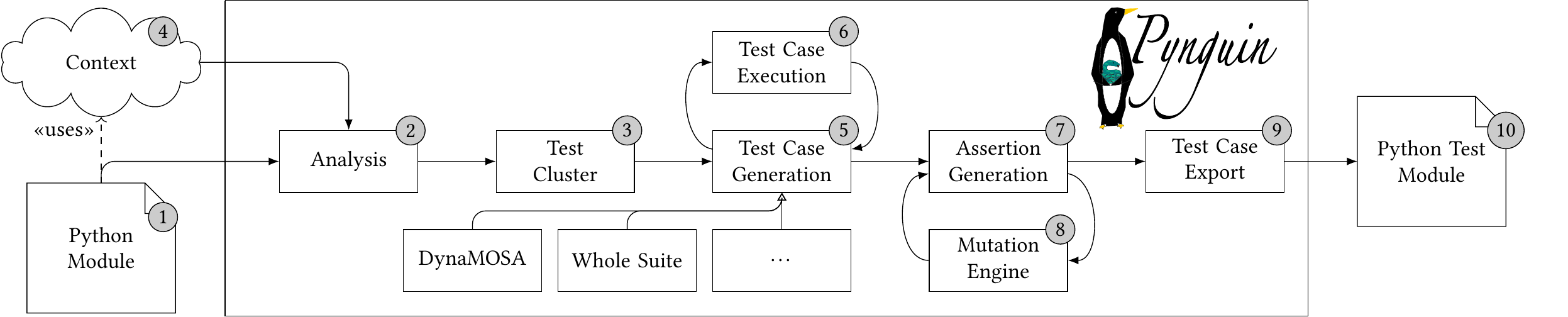}
  \Description{A flow-chart showing the steps of \pynguin's execution.
    It starts with a Python module that is analysed.
    Afterwards, different test generation algorithms generate input values
    in test cases;
    these test cases get executed to measure their fitness,
    before further test can be generated.
    The next stage generates assertions
    by iteratively mutating the code under test
    and executing it against the newly generated test cases.
    Finally, the generated test cases are exported into a Python test module.}
  \caption{\label{fig:pynguin-components}The execution steps of \pynguin.}
\end{figure*}

\Pynguin is an open-source framework written in and for
the Python programming language.
It uses search-based test generation to generate tests
that maximise code coverage.
\Pynguin by default incorporates type information
into the test-generation process.
However,
it is also able to generate covering test cases
for programs that do not explicitly provide type information.
Designed as an extendable framework,
\pynguin allows both researchers and practitioners
to explore established and new ideas
for test generation
in the context of a dynamically typed programming language,
such as further coverage criteria
or new test-generation algorithms.
%

Initial empirical evaluation~\cite{LKF20} of \pynguin shows
that the automated generation of regression tests is feasible
also for a dynamically typed programming language:
on average,
\pynguin achieved a branch coverage of up to \avgCoverage{}
on \numModules Python modules from \numProjects open-source libraries.
Our initial evaluation furthermore indicates
that type information is crucial
also for the test-generation process:
Incorporating type information
leads to significantly higher coverage levels;
our previous experiments show a median improvement of up to
\num[round-precision=2]{2.7} percentage points over all used projects,
depending on the test-generation algorithm~\cite{LKF21}.
This paper describes the inner workings of \pynguin
and how one can use and extend it.
%


\section{Test Generation with \pynguin}\label{sec:approach}


\Pynguin is written in Python
and requires at least Python~3.8 to run.
It can,
however,
generate unit tests also for Python projects
that are built for older versions of Python.
\Pynguin can be run as a standalone command-line application
or—which is recommended—inside a \toolname{Docker} container.
It is released under the GNU LGPL open-source licence.
%


\subsection{\pynguin's Components}\label{sec:approach-internal}

\Cref{fig:pynguin-components} shows the components of \pynguin 
and their interactions
throughout the test-generation process.
%

\Pynguin takes as input a Python module~(denoted
by \circled{1}
 in \cref{fig:pynguin-components}).\footnote{%
  Usually, a module in Python is equivalent to a source file.
  We currently restrict the support to modules written in Python
  because of the necessary code instrumentation.%
}
It then analyses the module to extract information~\circled{2}.
The extracted information consists,
among others,
of the declared classes, functions, and methods.
From this information
\pynguin builds the so-called \emph{test cluster}~\cite{WL05}~\circled{3}.
The test cluster contains all information about the module under test,
most importantly,
which classes, functions, and methods are declared,
and what their parameters are.
Furthermore,
\pynguin inspects the modules
that are transitively included by the module under test~(shown
as the \emph{context} in \cref{fig:pynguin-components}, \circled{4}).
From the context,
\pynguin extracts the types they define
by searching for those class definitions
that are available in the namespace of the module under test.
These types are then used as input-type candidates
during the test-generation phase.
\Pynguin selects classes, methods, and functions
from the test cluster during the generation
to build the test cases.
%

When constructing a test case,
\pynguin selects a function or method from the module under test.
Consider the example code snipped in \cref{lst:triangle}:
there is only one function in the module,
\lstinline[language=python]!triangle!,
which \pynguin selects as its target function.
It therefore adds a statement representing a method call to
\lstinline[language=python]!triangle!
to its internal test-case representation.
Afterwards,
\pynguin aims to fulfill the requirements of the function's parameters
in a backwards fashion.
In the example,
\pynguin knows from the type annotations
that \lstinline[language=python]!int! statements are required.
It therefore generates one to three variable assignment statements of the form
\lstinline[language=python]!var = <num>!
and adds them to the test case
before the function-call statement.
The number of \lstinline[language=python]!int! statements
as well as the generated values are chosen randomly by \pynguin,
because variable values can be used for more than one parameter.
\Cref{lst:test-triangle} shows two test cases
that have been created by this way.
In case a more complex object is required as a parameter,
\pynguin will attempt to generate it
by recursively fulfilling the parameters of the involved methods;
the necessary statements are also prepended
to the list of statements of the test case.
We provide a detailed example of this process in our previous work~\cite{LKF21}.
%

For test input generation~\circled{5}
the user can select between various well-established algorithms:
DynaMOSA~\cite{PKT18},
MIO~\cite{Arc17},
MOSA~\cite{PKT15},
random~\cite{PLE+07},
Whole Suite~\cite{FA13},
and Whole Suite with archive~\cite{RVA+17}.
Depending on the selected algorithm,
\pynguin generates one or many test cases.
It then executes the newly generated test cases
against the module under test
to measure the achieved coverage~\circled{6}.
Currently,
\pynguin can consider line or branch coverage as an optimisation goal
for its search algorithms.
To support other variants of coverage
one needs to provide further fitness functions;
further coverage criteria are planned for future work.
It is possible to select one sort of coverage for the optimisation
or a combination of many.
To measure coverage
we instrument Python's byte code on-the-fly to trace
which parts of the module under test have been executed by a generated test.
After evaluating fitness,
\pynguin continues with the next iteration of the test-generation algorithm.
This process stops
once a configurable stopping condition is satisfied, such as
a time limit or a predefined amount of algorithm iterations.
It is also possible to stop the generation
after all coverage goals have been met,
which means the generated tests achieve \SI{100}{\percent} coverage.
%

After the test-input generation
\pynguin optionally attempts to generate regression assertions~\cite{Xie06}
to not only execute the code under test
but also check its results~\circled{7}.
The approach implemented in \pynguin is based on mutation testing~\cite{FZ12}.
\Pynguin utilises a customised version of \MutPy~\cite{DH14}
to generate mutated versions from the original module under test~\circled{8}.
\MutPy executes the tests generated by the previous stage of \pynguin
against these mutants as well as the original module.
By tracing the values of object attributes and function returns,
\pynguin determines which values change on the mutated version,
compared to the original module.
For these values \pynguin generates assertions
that interpret the returned values on the original module
as the ground truth.
As a consequence,
the generated assertions are able to kill the aforementioned mutants
if they show different behaviour
compared to the original module.
An advantage of generating regression tests this way is
that it implicitly minimises the number of assertions
present in the resulting test cases.
Finally, \pynguin generates Python source code~\circled{9}
from its internal representation of the test cases
and exports the source code
in the style of the popular
\toolname{PyTest}\footnote{%
  \url{https://pytest.org}, last access 2022–02–10.%
} framework into a Python module~\circled{10}.
Further styles,
for example,
\texttt{unittest} from Python's standard API,
can also be integrated easily.
%

Each of the stages of \pynguin
is built as modular and as independent of the others
as possible.
\Pynguin itself is furthermore built with extendability in mind.
This allows to replace stages and components easily.
%


\subsection{Using \pynguin}\label{sec:approach-usage}

\Pynguin is written in Python.
It can thus be either executed after checking out its source code
or—more conveniently—be installed
from the Python Package Index~(PyPI)\footnote{%
  \url{https://pypi.org/project/pynguin}, last accessed 2022–02–10.%
} via the \texttt{pip} utility tool.
%

The primary usage of \pynguin is as a command-line application.
It also provides a rudimentary API
that allows controlling the framework from inside another application
without the need to launch an external process.
As future work
we plan to enhance \pynguin's public API
such that it can also be used as a library for test generation
within other projects
without the necessity to execute it as a standalone application.
Our presentation here,
however,
will only discuss the command-line interface.
We refer the interested reader to \pynguin's documentation\footnote{%
  \url{https://pynguin.readthedocs.io}, last accessed 2022–02–10.%
}, which describes the API.
One can get an overview of all command-line arguments
using the \verb!--help! option
after installing \pynguin:
\begin{lstlisting}[language=bash]
$ pynguin --help
\end{lstlisting}

Please note that \pynguin requires the user
to set the environment variable \verb!PYNGUIN_DANGER_AWARE!;
\pynguin executes the code under test
with arbitrary random inputs.
Depending on the code under test
this can cause side effects and harm to the user's system.
By setting the environment variable to an arbitrary value
the user confirms to \pynguin that they are aware of this risk.
%

The main arguments of \pynguin are \verb!--project-path!
to specify the path of the project
\pynguin should generate tests for,
\verb!--module-name! to specify the name of the module
to generate tests for,
and the \verb!--output-path!,
where \pynguin stores the generated test cases.
\Pynguin requires the user to set at least those three parameters;
all further arguments come with documented default values
yielded by the \verb!--help! parameter.
%

Consider the example in \cref{lst:triangle},
saved to a module \texttt{triangle.py} in the current work directory.
One can now run \pynguin with minimal configuration options:
\begin{lstlisting}[language=bash]
$ pynguin \
    --project-path ./ \
    --output-path /tmp/pynguin-tests \
    --module-name triangle
\end{lstlisting}
This results in \pynguin generating test cases using DynaMOSA~(the current
default algorithm).
It stores them in files to the folder \texttt{/tmp/pynguin-tests}.
Now suppose that the user wants to generate tests for the same module
but with the MIO algorithm instead of DynaMOSA.
All they need to do is to add \verb!--algorithm MIO! to their command line:
\begin{lstlisting}[language=bash]
$ pynguin \
    --project-path ./ \
    --output-path /tmp/pynguin-results \
    --module-name triangle \
    --algorithm MIO
\end{lstlisting}
Similarly,
one can set further configuration options.
\Cref{lst:test-triangle} shows an excerpt of the generated result.\footnote{%
  Since the generation process is based on random numbers,
  the exact result might differ.
  It still should look similar.
}
The two shown test cases in the result
execute the \lstinline[language=python]!triangle! function
with different parameter values
to execute different branches of the function's implementation;
the test cases also provide assertions
that check on the returned value of the \lstinline[language=python]!triange!
function.
Please note that re-executing \pynguin will overwrite the resulting files.

\begin{lstlisting}[%
  float=t,%
  language=Python,%
  caption={A simple function checking for triangle properties.},%
  label={lst:triangle},%
]
def triangle(x: int, y: int, z: int) -> str:
  if x == y == z:
    return "Equilateral triangle"
  elif x == y or y == z or x == z:
    return "Isosceles triangle"
  else:
    return "Scalene triangle"
\end{lstlisting}

\begin{lstlisting}[%
  float=t,%
  language=Python,%
  caption={An excerpt of the test cases generated by \pynguin.},%
  label={lst:test-triangle},%
]
import example as module0

def test_case_0():
    int_0 = 4271
    int_1 = -3706
    str_0 = module0.triangle(int_0, int_0, int_1)
    assert str_0 == "Isosceles triangle"

def test_case_1():
    int_0 = -163
    int_1 = 484
    int_2 = 77
    str_0 = module0.triangle(int_0, int_1, int_2)
    assert str_0 == "Scalene triangle"
\end{lstlisting}

In the aforementioned settings,
\pynguin will not print any output to the terminal.
A more verbose output can be achieved by adding the \texttt{-v}
or \texttt{-vv} parameter.
%


\subsection{Dynamic Typing}\label{sec:approach-typing}

As Python is a dynamically typed language,
it does not require the user to specify any type information,
although recent versions of the language support annotations for such information.
\Pynguin aims to parse type annotations
from the source code if they are available~(configurable as a parameter
to \pynguin).
Any parsed information about parameter types of functions and methods
as well about their return types is incorporated into the test cluster.
The type information stored in the test cluster
allows \pynguin to select specific objects
to satisfy the requirements of the parameters
when generating test cases for a specific function or method.
%

Besides extracting type information from annotations in the source code,
\pynguin can also be extended to query external type-inference tools.
If no type annotations are available for the code under test
\pynguin considers all available types from the test cluster
as candidates during input generation.
In this case,
\pynguin currently selects one of the available types
from the test cluster randomly.
%
%

Let us again consider the triangle-classification function
from \cref{lst:triangle}.
Suppose there were no type annotation present:
As a consequence,
\pynguin could only guess the types of the parameters
and might come up with objects of arbitrary types
available from the test cluster.
Since our example does neither define new types
nor import any modules,
only the so called \enquote{builtins} are available:
basic types such as \lstinline[language=python]!float!
or \lstinline[language=python]!str!.
Choosing parameter values of type \lstinline[language=python]!float! instead of
\lstinline[language=python]!int! would be a reasonable choice for this triangle
function.
However,
the following test case would also be valid for the Python language if
there is no type information available:
\begin{lstlisting}[language=python]
import example as module0

def test_case_2():
    list_0 = ["foo", "bar"]
    str_0 = module0.triangle(list_0, list_0, list_0)
    assert str_0 == "Equilateral triangle"
\end{lstlisting}

Checking the triangle type for a list of strings
is not a reasonable thing,
although the language would permit it.
Note that the above test case is valid
because it executes the program without crashes
and covers parts of the code of the \lstinline[language=python]!triangle!
function;
it therefore contributes towards the optimisation goal of high coverage.
However,
the test case also shows that coverage may not be a good metric
for the effects of types.
Using an unexpected type as an input
may often also simply lead to crashes of the program under test,
for example,
if the code attempts to access non-existing attributes.
Furthermore,
the lack of type information can also prevent \pynguin
from being able to instantiate the correct objects.
This simple example shows
that type information is crucial for test generation
to generate not only covering
but also valid and useful tests.
While we strongly advocate the usage of type annotations
to improve \pynguin's resulting test cases,
the use of alternative means such as type inference
is an open research problem.
Enabling future research to address this problem
is a core motivation for building \pynguin.
%


\subsection{Extending \pynguin}\label{sec:approach-extend}

\Pynguin is a framework
that allows easy extension at various points.
Many extensions can be built by implementing
only few classes.
For example, adding a new test-generation algorithm
can be achieved by extending the abstract class
\sourcecode{TestGenerationStrategy}
and implementing its \sourcecode{generate\_tests} method.
It is also necessary to register the new algorithm
in the configuration and the algorithm instantiation.
The algorithm instantiation allows providing predefined operators
to the new algorithm
such as a factory for new chromosomes
or fitness functions for the search objectives.
We designed these components with evolutionary algorithms in mind;
one can of course also define purely random-based algorithms
as we have done with our Random algorithm,
based on the \toolname{Randoop} algorithm~\cite{PLE+07}.
\begin{lstlisting}[%
  float=t,%
  language=Python,%
  caption={A random sampling strategy for test input generation.},%
  label={lst:randomsampling},%
]
class RandomSamplingStrategy(TestGenerationStrategy):
  def generate_tests(self):
    self.before_search_start()
    solution = self._chromosome_factory.get_chromosome()
    self._archive.update([solution])
    test_suite = self.create_test_suite(
        self._archive.solutions
    )
    self.before_first_search_iteration(solution)

    while (
        self.resources_left()
        and test_suite.get_fitness() != 0.0
    ):
      new = self._chromosome_factory.get_chromosome()
      self._archive.update([new])
      test_suite = self.create_test_suite(
          self._archive.solutions
      )
      self.after_search_iteration(test_suite)

    self.after_search_finish()
    return self.create_test_suite(
        self._archive.solutions
    )
\end{lstlisting}
%

To demonstrate how simple this can be,
\cref{lst:randomsampling} implements an algorithm
that is based on random test-case sampling:
it generates a random test case by adding a method call
and its dependencies—the object to call the method on
as well as objects to fill the parameter values~(see Sec.~2 of our
previous work~\cite{LKF21} for a detailed illustration of this process).
Our example algorithm utilises the archive
that is also used for the MOSA~\cite{PKT15} algorithm
to store generated test cases
that cover certain coverage goals in order to keep track of them.
Now the algorithm loops until the stopping condition is fulfilled
or the test cases stored in the archive
cover all coverage goals.
In each loop iteration,
the algorithm randomly samples another test case.
It stores the new test case in the archive as well.
The given implementation also calls some helper methods
that keep track of the test-generation process;
for example,
they track the achieved coverage value
over the generation time.
In the end,
the example algorithm creates a test suite,
that is, a collection of the generated test cases,
and returns them back to the framework.
We already implemented this algorithm in \pynguin;
one can select it by setting
\verb!--algorithm RANDOM_TEST_CASE_SEARCH!.
%

Similarly,
other parts of \pynguin can be extended.
Examples are test-case export in different styles
by implementing an AST visitor
or the incorporation of type-inference techniques
by querying external type-inference tools.
%


\section{Evaluation}\label{sec:evaluation}

We evaluated \pynguin
by conducting a small experiment for this paper,
using the aforementioned algorithms for test generation.
We used \numModules modules from our previous work~\cite{LKF21}
for our evaluation
and ran \pynguin in version~0.17.0~\cite{Luk22Pynguin0170}
\numIterations times on each module
and in each configuration
to minimise the influence of randomness.
For the experiment,
we set the timeout for test generation to \timeout.
%
In this work we only give few insights;
for a more extensive evaluation
we refer the reader to our previous work~\cite{LKF20,LKF21},
which not only studies the differences
between various algorithms in greater detail
but also investigates on the influence of type information.

\begin{figure}[t]
  \centering
  \includegraphics[width=0.95\linewidth]{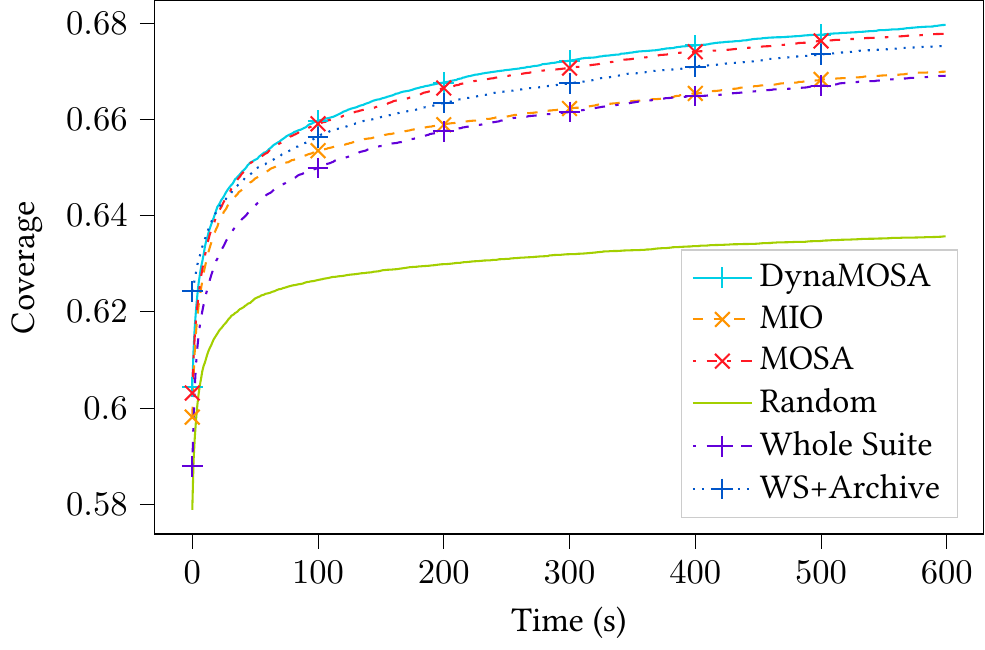}
  \Description{A line plot showing the coverage results over time for the
    six algorithms.
    DynaMOSA performs best, at around \avgCoverage branches covered,
    followed by MOSA, Whole Suite with Archive, MIO, and Whole Suite.
    The Random algorithm yields the lowest coverage results.
  }
  \caption{\label{fig:coverage-over-time}Development of the coverage over time.}
\end{figure}

To gain insights on the performance of the different algorithms,
we measured branch coverage.
\Cref{fig:coverage-over-time} shows the development
of the mean coverage per configuration
over the generation time of \timeout.
One can clearly see that the search-based techniques
outperform the random algorithm.
The five search-based algorithms,
however,
only show small differences,
with DynaMOSA achieving the highest
and Whole Suite the lowest coverage values~(mean branch coverage for
DynaMOSA\@: \avgCoverageDynaMOSA,
MIO\@: \avgCoverageMIO,
MOSA\@: \avgCoverageMOSA,
Random\@: \avgCoverageRandom,
Whole Suite\@: \avgCoverageWS,
and for Whole Suite with archive\@: \avgCoverageWSA).
These results are in line with previous research~\cite{CAF+18,PKT18a}
in the context of statically typed languages:
the search-based algorithms achieve a higher average coverage
than the random algorithm.
Furthermore, DynaMOSA yields the highest coverage values.
%


\section{Conclusions}\label{sec:conclusions}

The increasing popularity of Python
requires the availability of many types of tools
to aid the developers.
We introduce \pynguin,
an automated unit test generation framework,
designed to support developers
when implementing unit tests manually.
\Pynguin is available as a command-line application,
which is the de-facto standard for many developer-aiding tools.
While this provides great flexibility for users,
\pynguin's modular design also enables further extensions
and research in the field of automated unit test generation
for dynamically typed languages.
In this work we have summarised the features of \pynguin.
By providing \pynguin as open source,
we hope to foster further research
as well as its applicability in practice.
Further information on \pynguin,
its documentation,
and source code
are available at
\begin{center}
  \url{https://www.pynguin.eu}
\end{center}


\begin{acks}
  This work is supported by
  \grantsponsor{FR~2955/4-1}{DFG}{https://gepris.dfg.de/gepris/projekt/434705464}
  project~\grantnum{DFG}{FR~2955/4-1}.
\end{acks}

\bibliographystyle{ACM-Reference-Format}
\bibliography{related}


\begin{thebibliography}{22}


\ifx \showCODEN    \undefined \def \showCODEN     #1{\unskip}     \fi
\ifx \showDOI      \undefined \def \showDOI       #1{#1}\fi
\ifx \showISBNx    \undefined \def \showISBNx     #1{\unskip}     \fi
\ifx \showISBNxiii \undefined \def \showISBNxiii  #1{\unskip}     \fi
\ifx \showISSN     \undefined \def \showISSN      #1{\unskip}     \fi
\ifx \showLCCN     \undefined \def \showLCCN      #1{\unskip}     \fi
\ifx \shownote     \undefined \def \shownote      #1{#1}          \fi
\ifx \showarticletitle \undefined \def \showarticletitle #1{#1}   \fi
\ifx \showURL      \undefined \def \showURL       {\relax}        \fi
\providecommand\bibfield[2]{#2}
\providecommand\bibinfo[2]{#2}
\providecommand\natexlab[1]{#1}
\providecommand\showeprint[2][]{arXiv:#2}

\bibitem[Arcuri(2017)]%
        {Arc17}
\bibfield{author}{\bibinfo{person}{Andrea Arcuri}.}
  \bibinfo{year}{2017}\natexlab{}.
\newblock \showarticletitle{Many Independent Objective {(MIO)} Algorithm for
  Test Suite Generation}. In \bibinfo{booktitle}{\emph{International Symposium
  on Search Based Software Engineering~(SSBSE)}}
  \emph{(\bibinfo{series}{Lecture Notes in Computer Science},
  Vol.~\bibinfo{volume}{10452})}. \bibinfo{publisher}{Springer},
  \bibinfo{pages}{3--17}.
\newblock
\urldef\tempurl%
\url{https://doi.org/10.1007/978-3-319-66299-2\_1}
\showDOI{\tempurl}


\bibitem[Campos et~al\mbox{.}(2018)]%
        {CAF+18}
\bibfield{author}{\bibinfo{person}{José Campos}, \bibinfo{person}{Yan Ge},
  \bibinfo{person}{Nasser Albunian}, \bibinfo{person}{Gordon Fraser},
  \bibinfo{person}{Marcelo Eler}, {and} \bibinfo{person}{Andrea Arcuri}.}
  \bibinfo{year}{2018}\natexlab{}.
\newblock \showarticletitle{An empirical evaluation of evolutionary algorithms
  for unit test suite generation}.
\newblock \bibinfo{journal}{\emph{Information {\&} Software Technology}}
  \bibinfo{volume}{104} (\bibinfo{year}{2018}), \bibinfo{pages}{207--235}.
\newblock
\urldef\tempurl%
\url{https://doi.org/10.1016/j.infsof.2018.08.010}
\showDOI{\tempurl}


\bibitem[Derezinska and Ha{\l}as(2014)]%
        {DH14}
\bibfield{author}{\bibinfo{person}{Anna Derezinska} {and}
  \bibinfo{person}{Konrad Ha{\l}as}.} \bibinfo{year}{2014}\natexlab{}.
\newblock \showarticletitle{Experimental Evaluation of Mutation Testing
  Approaches to Python Programs}. In \bibinfo{booktitle}{\emph{International
  Conference on Software Testing, Verification and Validation
  Workshops~(ICST-Workshops)}}. \bibinfo{publisher}{{IEEE} Computer Society},
  \bibinfo{pages}{156--164}.
\newblock
\urldef\tempurl%
\url{https://doi.org/10.1109/ICSTW.2014.24}
\showDOI{\tempurl}


\bibitem[Duran and Ntafos(1984)]%
        {DN84}
\bibfield{author}{\bibinfo{person}{Joe~W. Duran} {and}
  \bibinfo{person}{Simeon~C. Ntafos}.} \bibinfo{year}{1984}\natexlab{}.
\newblock \showarticletitle{An Evaluation of Random Testing}.
\newblock \bibinfo{journal}{\emph{{IEEE} Transactions on Software Engineering}}
  \bibinfo{volume}{10}, \bibinfo{number}{4} (\bibinfo{year}{1984}),
  \bibinfo{pages}{438--444}.
\newblock
\urldef\tempurl%
\url{https://doi.org/10.1109/TSE.1984.5010257}
\showDOI{\tempurl}


\bibitem[Fraser and Arcuri(2011)]%
        {FA11}
\bibfield{author}{\bibinfo{person}{Gordon Fraser} {and} \bibinfo{person}{Andrea
  Arcuri}.} \bibinfo{year}{2011}\natexlab{}.
\newblock \showarticletitle{EvoSuite: Automatic Test Suite Generation for
  Object-Oriented Software}. In \bibinfo{booktitle}{\emph{Joint Meeting of the
  European Software Engineering Conference and the Symposium on the Foundations
  of Software Engineering~(ESEC/FSE)}}. \bibinfo{publisher}{{ACM}},
  \bibinfo{pages}{416--419}.
\newblock
\urldef\tempurl%
\url{https://doi.org/10.1145/2025113.2025179}
\showDOI{\tempurl}


\bibitem[Fraser and Arcuri(2013)]%
        {FA13}
\bibfield{author}{\bibinfo{person}{Gordon Fraser} {and} \bibinfo{person}{Andrea
  Arcuri}.} \bibinfo{year}{2013}\natexlab{}.
\newblock \showarticletitle{Whole Test Suite Generation}.
\newblock \bibinfo{journal}{\emph{{IEEE} Transactions on Software Engineering}}
  \bibinfo{volume}{39}, \bibinfo{number}{2} (\bibinfo{year}{2013}),
  \bibinfo{pages}{276--291}.
\newblock
\urldef\tempurl%
\url{https://doi.org/10.1109/TSE.2012.14}
\showDOI{\tempurl}


\bibitem[Fraser and Zeller(2012)]%
        {FZ12}
\bibfield{author}{\bibinfo{person}{Gordon Fraser} {and}
  \bibinfo{person}{Andreas Zeller}.} \bibinfo{year}{2012}\natexlab{}.
\newblock \showarticletitle{Mutation-Driven Generation of Unit Tests and
  Oracles}.
\newblock \bibinfo{journal}{\emph{{IEEE} Transactions on Software Engineering}}
  \bibinfo{volume}{38}, \bibinfo{number}{2} (\bibinfo{year}{2012}),
  \bibinfo{pages}{278--292}.
\newblock
\urldef\tempurl%
\url{https://doi.org/10.1109/TSE.2011.93}
\showDOI{\tempurl}


\bibitem[Gong et~al\mbox{.}(2015)]%
        {GPS+15}
\bibfield{author}{\bibinfo{person}{Liang Gong}, \bibinfo{person}{Michael
  Pradel}, \bibinfo{person}{Manu Sridharan}, {and} \bibinfo{person}{Koushik
  Sen}.} \bibinfo{year}{2015}\natexlab{}.
\newblock \showarticletitle{{DLint:} Dynamically Checking Bad Coding Practices
  in {JavaScript}}. In \bibinfo{booktitle}{\emph{International Symposium on
  Software Testing and Analysis~(ISSTA)}}. \bibinfo{publisher}{{ACM}},
  \bibinfo{pages}{94--105}.
\newblock
\urldef\tempurl%
\url{https://doi.org/10.1145/2771783.2771809}
\showDOI{\tempurl}


\bibitem[Holmes et~al\mbox{.}(2018)]%
        {HGP+18}
\bibfield{author}{\bibinfo{person}{Josie Holmes}, \bibinfo{person}{Alex Groce},
  \bibinfo{person}{Jervis Pinto}, \bibinfo{person}{Pranjal Mittal},
  \bibinfo{person}{Pooria Azimi}, \bibinfo{person}{Kevin Kellar}, {and}
  \bibinfo{person}{James O'Brien}.} \bibinfo{year}{2018}\natexlab{}.
\newblock \showarticletitle{{TSTL:} the template scripting testing language}.
\newblock \bibinfo{journal}{\emph{International Journal on Software Tools for
  Technology Transfer}} \bibinfo{volume}{20}, \bibinfo{number}{1}
  (\bibinfo{year}{2018}), \bibinfo{pages}{57--78}.
\newblock
\urldef\tempurl%
\url{https://doi.org/10.1007/s10009-016-0445-y}
\showDOI{\tempurl}


\bibitem[Lukasczyk et~al\mbox{.}(2020)]%
        {LKF20}
\bibfield{author}{\bibinfo{person}{Stephan Lukasczyk}, \bibinfo{person}{Florian
  Kroi\ss{}}, {and} \bibinfo{person}{Gordon Fraser}.}
  \bibinfo{year}{2020}\natexlab{}.
\newblock \showarticletitle{Automated Unit Test Generation for Python}. In
  \bibinfo{booktitle}{\emph{International Symposium on Search Based Software
  Engineering~(SSBSE)}} \emph{(\bibinfo{series}{Lecture Notes in Computer
  Science}, Vol.~\bibinfo{volume}{12420})}. \bibinfo{publisher}{Springer},
  \bibinfo{pages}{9--24}.
\newblock
\urldef\tempurl%
\url{https://doi.org/10.1007/978-3-030-59762-7\_2}
\showDOI{\tempurl}


\bibitem[Lukasczyk et~al\mbox{.}(2021)]%
        {LKF21}
\bibfield{author}{\bibinfo{person}{Stephan Lukasczyk}, \bibinfo{person}{Florian
  Kroiß}, {and} \bibinfo{person}{Gordon Fraser}.}
  \bibinfo{year}{2021}\natexlab{}.
\newblock \showarticletitle{An Empirical Study of Automated Unit Test
  Generation for Python}.
\newblock \bibinfo{journal}{\emph{CoRR}}  \bibinfo{volume}{abs/2111.05003}
  (\bibinfo{year}{2021}).
\newblock
\showeprint[arXiv]{2111.05003}


\bibitem[Lukasczyk et~al\mbox{.}(2022)]%
        {Luk22Pynguin0170}
\bibfield{author}{\bibinfo{person}{Stephan Lukasczyk}, \bibinfo{person}{Florian
  Kroiß}, \bibinfo{person}{Gordon Fraser}, {and} \bibinfo{person}{Pynguin
  Contributors}.} \bibinfo{year}{2022}\natexlab{}.
\newblock \bibinfo{booktitle}{\emph{se2p/pynguin: Pynguin v0.17.0}}.
\newblock
\urldef\tempurl%
\url{https://doi.org/10.5281/zenodo.5971799}
\showDOI{\tempurl}


\bibitem[MacIver and Hatfield{-}Dodds(2019)]%
        {MH19}
\bibfield{author}{\bibinfo{person}{David MacIver} {and} \bibinfo{person}{Zac
  Hatfield{-}Dodds}.} \bibinfo{year}{2019}\natexlab{}.
\newblock \showarticletitle{Hypothesis: {A} new approach to property-based
  testing}.
\newblock \bibinfo{journal}{\emph{Journal of Open Source Software}}
  \bibinfo{volume}{4}, \bibinfo{number}{43} (\bibinfo{year}{2019}),
  \bibinfo{pages}{1891}.
\newblock
\urldef\tempurl%
\url{https://doi.org/10.21105/joss.01891}
\showDOI{\tempurl}


\bibitem[McMinn(2004)]%
        {McM04}
\bibfield{author}{\bibinfo{person}{Phil McMinn}.}
  \bibinfo{year}{2004}\natexlab{}.
\newblock \showarticletitle{Search-based Software Test Data Generation: A
  Survey}.
\newblock \bibinfo{journal}{\emph{Journal of Software Testing, Verification and
  Reliability}} \bibinfo{volume}{14}, \bibinfo{number}{2}
  (\bibinfo{year}{2004}), \bibinfo{pages}{105--156}.
\newblock
\urldef\tempurl%
\url{https://doi.org/10.1002/stvr.294}
\showDOI{\tempurl}


\bibitem[Pacheco et~al\mbox{.}(2007)]%
        {PLE+07}
\bibfield{author}{\bibinfo{person}{Carlos Pacheco},
  \bibinfo{person}{Shuvendu~K. Lahiri}, \bibinfo{person}{Michael~D. Ernst},
  {and} \bibinfo{person}{Thomas Ball}.} \bibinfo{year}{2007}\natexlab{}.
\newblock \showarticletitle{Feedback-Directed Random Test Generation}. In
  \bibinfo{booktitle}{\emph{International Conference on Software
  Engineering~(ICSE)}}. \bibinfo{publisher}{{IEEE} Computer Society},
  \bibinfo{pages}{75--84}.
\newblock
\urldef\tempurl%
\url{https://doi.org/10.1109/ICSE.2007.37}
\showDOI{\tempurl}


\bibitem[Panichella et~al\mbox{.}(2015)]%
        {PKT15}
\bibfield{author}{\bibinfo{person}{Annibale Panichella},
  \bibinfo{person}{Fitsum~Meshesha Kifetew}, {and} \bibinfo{person}{Paolo
  Tonella}.} \bibinfo{year}{2015}\natexlab{}.
\newblock \showarticletitle{Reformulating Branch Coverage as a Many-Objective
  Optimization Problem}. In \bibinfo{booktitle}{\emph{International Conference
  on Software Testing, Verification and Validation~(ICST)}}.
  \bibinfo{publisher}{{IEEE} Computer Society}, \bibinfo{pages}{1--10}.
\newblock
\urldef\tempurl%
\url{https://doi.org/10.1109/ICST.2015.7102604}
\showDOI{\tempurl}


\bibitem[Panichella et~al\mbox{.}(2018a)]%
        {PKT18}
\bibfield{author}{\bibinfo{person}{Annibale Panichella},
  \bibinfo{person}{Fitsum~Meshesha Kifetew}, {and} \bibinfo{person}{Paolo
  Tonella}.} \bibinfo{year}{2018}\natexlab{a}.
\newblock \showarticletitle{Automated Test Case Generation as a Many-Objective
  Optimisation Problem with Dynamic Selection of the Targets}.
\newblock \bibinfo{journal}{\emph{{IEEE} Transactions on Software Engineering}}
  \bibinfo{volume}{44}, \bibinfo{number}{2} (\bibinfo{year}{2018}),
  \bibinfo{pages}{122--158}.
\newblock
\urldef\tempurl%
\url{https://doi.org/10.1109/TSE.2017.2663435}
\showDOI{\tempurl}


\bibitem[Panichella et~al\mbox{.}(2018b)]%
        {PKT18a}
\bibfield{author}{\bibinfo{person}{Annibale Panichella},
  \bibinfo{person}{Fitsum~Meshesha Kifetew}, {and} \bibinfo{person}{Paolo
  Tonella}.} \bibinfo{year}{2018}\natexlab{b}.
\newblock \showarticletitle{A large scale empirical comparison of
  state-of-the-art search-based test case generators}.
\newblock \bibinfo{journal}{\emph{Information {\&} Software Technology}}
  \bibinfo{volume}{104} (\bibinfo{year}{2018}), \bibinfo{pages}{236--256}.
\newblock
\urldef\tempurl%
\url{https://doi.org/10.1016/j.infsof.2018.08.009}
\showDOI{\tempurl}


\bibitem[Pradel and Sen(2015)]%
        {PS15}
\bibfield{author}{\bibinfo{person}{Michael Pradel} {and}
  \bibinfo{person}{Koushik Sen}.} \bibinfo{year}{2015}\natexlab{}.
\newblock \showarticletitle{The Good, the Bad, and the Ugly: An Empirical Study
  on Implicit Type Conversions in JavaScript}. In
  \bibinfo{booktitle}{\emph{European Conference on Object-Oriented
  Programming~(ECOOP)}} \emph{(\bibinfo{series}{Leibnitz International
  Proceedings in Informatics~(LIPIcs)}, Vol.~\bibinfo{volume}{37})}.
  \bibinfo{publisher}{Schloss Dagstuhl – Leibnitz-Zentrum für Informatik},
  \bibinfo{pages}{519--541}.
\newblock
\urldef\tempurl%
\url{https://doi.org/10.4230/LIPIcs.ECOOP.2015.519}
\showDOI{\tempurl}


\bibitem[Rojas et~al\mbox{.}(2017)]%
        {RVA+17}
\bibfield{author}{\bibinfo{person}{José~Miguel Rojas}, \bibinfo{person}{Mattia
  Vivanti}, \bibinfo{person}{Andrea Arcuri}, {and} \bibinfo{person}{Gordon
  Fraser}.} \bibinfo{year}{2017}\natexlab{}.
\newblock \showarticletitle{A detailed investigation of the effectiveness of
  whole test suite generation}.
\newblock \bibinfo{journal}{\emph{Empirical Software Engineering}}
  \bibinfo{volume}{22}, \bibinfo{number}{2} (\bibinfo{year}{2017}),
  \bibinfo{pages}{852--893}.
\newblock
\urldef\tempurl%
\url{https://doi.org/10.1007/s10664-015-9424-2}
\showDOI{\tempurl}


\bibitem[Wappler and Lammermann(2005)]%
        {WL05}
\bibfield{author}{\bibinfo{person}{Stefan Wappler} {and} \bibinfo{person}{Frank
  Lammermann}.} \bibinfo{year}{2005}\natexlab{}.
\newblock \showarticletitle{Using evolutionary algorithms for the unit testing
  of object-oriented software}. In \bibinfo{booktitle}{\emph{Annual Conference
  on Genetic and Evolutionary Computation~(GECCO)}}.
  \bibinfo{pages}{1053--1060}.
\newblock
\urldef\tempurl%
\url{https://doi.org/10.1145/1068009.1068187}
\showDOI{\tempurl}


\bibitem[Xie(2006)]%
        {Xie06}
\bibfield{author}{\bibinfo{person}{Tao Xie}.} \bibinfo{year}{2006}\natexlab{}.
\newblock \showarticletitle{Augmenting Automatically Generated Unit-Test Suites
  with Regression Oracle Checking}. In \bibinfo{booktitle}{\emph{European
  Conference on Object-Oriented Programming~(ECOOP)}}
  \emph{(\bibinfo{series}{Lecture Notes in Computer Science},
  Vol.~\bibinfo{volume}{4067})}. \bibinfo{publisher}{Springer},
  \bibinfo{pages}{380--403}.
\newblock
\urldef\tempurl%
\url{https://doi.org/10.1007/11785477\_23}
\showDOI{\tempurl}


\end{thebibliography}

\end{document}